\documentclass[a4paper,fleqn]{cas-dc}
\renewcommand{\printorcid}{}

\usepackage{apacite}
\usepackage[round]{natbib} 

\usepackage{breakcites}

\usepackage{float}
\usepackage[section]{placeins}

\restylefloat{table}
\usepackage{diagbox}
\usepackage{adjustbox}

\def\tsc#1{\csdef{#1}{\textsc{\lowercase{#1}}\xspace}}
\tsc{WGM}
\tsc{QE}
\tsc{EP}
\tsc{PMS}

\tsc{BEC}
\tsc{DE}

\begin{document}
\let\WriteBookmarks\relax
\def\floatpagepagefraction{1}
\def\textpagefraction{.001}
\shorttitle{Macroeconomic forecasting through news, emotions and narrative}
\shortauthors{Sonja Tilly et~al.}

\pagenumbering{arabic}

\title [mode = title]{Macroeconomic forecasting through news, emotions and narrative}                      

\author[1]{Sonja Tilly}

\cormark[1]

\ead{sonja.tilly.19@ucl.ac.uk}

\cortext[cor1]{Corresponding author}

\address[1]{UCL, Computer Science Dep, 66 - 72 Gower St, Bloomsbury, WC1E 6EA London, UK}

\author[2]{Markus Ebner}
\address[2]{Quoniam Asset Management, Westhafen Tower, Westhafenplatz 1, 60327 Frankfurt am Main, Germany}
\ead{markus.ebner@quoniam.com}

\author[1,3]{Giacomo Livan}
\address[3]{Systemic Risk Centre, London School of Economics and Political Science, London, WC2A 2AE, UK}
\ead{g.livan@ucl.ac.uk}

\nonumnote{Abbreviations. GDELT: Global Database of Events, Language and Tone; GKG: Global Knowledge Graph; GCAM: Content Analysis Measure Systems; CNT: Conviction Narrative Theory; Bi-LSTM: bi-directional long short term memory neural network; RNN: recurrent neural network; IP: industrial production; CPI: consumer price index; PLS: partial least squares}

\begin{abstract}
This study proposes a new method of incorporating emotions from newspaper articles into macroeconomic forecasts, attempting to forecast industrial production and consumer prices leveraging narrative and sentiment from global newspapers. For the most part, existing research includes positive and negative tone only to improve macroeconomic forecasts, focusing predominantly on large economies such as the US. These works use mainly anglophone sources of narrative, thus not capturing the entire complexity of the multitude of emotions contained in global news articles. This study expands the existing body of research by incorporating a wide array of emotions from newspapers around the world -- extracted from the Global Database of Events, Language and Tone (GDELT) -- into macroeconomic forecasts. We present a thematic data filtering methodology based on a bi-directional long short term memory neural network (Bi-LSTM) for extracting emotion scores from GDELT and demonstrate its effectiveness by comparing results for filtered and unfiltered data. We model industrial production and consumer prices across a diverse range of economies using an autoregressive framework, and find that including emotions from global newspapers significantly improves forecasts compared to three autoregressive benchmark models. We complement our forecasts with an interpretability analysis on distinct groups of emotions and find that emotions associated with happiness and anger have the strongest predictive power for the variables we predict. 

\end{abstract}


\maketitle

\section{Introduction}

Recent developments in automated language analysis have allowed to quantify the elusive yet intuitive notion of narrative, and to quantify its predictive power in relation to changes in social systems.

Research in psychology and cognitive sciences has examined the role emotions and narrative play in decision making and judgement \citep{brosch2013impact, clore2009affective, bruner1990acts}. These studies show that emotions can help individuals make decisions in complex scenarios with uncertain outcomes. 
Keynes uses the term ``animal spirits'' to describe the dispositions and emotions that drive human actions, with the results of this behaviour measurable in terms of economic indices such as consumer confidence \citep{keynes2018general}. Shiller finds that unsettling narrative led to events such as the Great Depression in the 1920s and the Global Financial Crisis in 2008/9, arguing that narrative is a means of predicting the economy \citep{shiller2017narrative}. A recent theoretical development -- known as Conviction Narrative Theory (CNT) -- draws on the concept that to be sufficiently confident to act, agents create narratives supporting their expectations of the outcome of their actions \citep{nyman2018news}. For instance, a study on CNT tracks changes in narrative and shows that they precede changes in economic growth \citep{tuckett2014bringing}. 

Media is an established, multi-functional tool for governments, corporations and individuals to disseminate information, connect and interact. As such, it is a major conduit for news narrative. Nowadays, most forms of media have an online presence and produce huge volumes of data. This data contains information in the form of opinions and sentiment about financial markets and the economy, which may not yet be reflected in macroeconomic variables.

Over recent years, researchers have explored sentiment from different types of media and its usefulness for the prediction of the economy and financial markets. Studies examine how to process large amounts of unstructured data from a variety of sources in order to extract signals \citep{buonoevaluation, elshendy2018using}. Other works outline approaches to incorporate such signals into a predictive model, for instance to improve the monitoring of the economy and financial forecasting \citep{levenberg2014predicting, slaper2018digital}.

Media sentiment prediction has a wide range of application domains that Rousidis et al group into finance, marketing and sociopolitical \citep{rousidis2020social}. Within the finance domain, studies explore media sentiment prediction either for specific assets or markets (micro level) \citep{allen2019daily} or for different aspects of the economy (macro level) \citep{ardia2019questioning}.

Existing research incorporates largely positive and negative tone to improve macroeconomic forecasts, thus not capturing the entire complexity of the multitude of emotions contained in global news articles. Most works use anglophone sources of narrative, focusing predominantly on large economies such as the US.

This study advances the existing body of research by incorporating a wide array of emotions from newspapers around the world into macroeconomic forecasts using data from the Global Database of Events, Language and Tone (GDELT) \citep{leetaru_the_nodate}. GDELT is a research collaboration that analyses global news articles and extracts items such as themes, emotions, locations, and many more. We employ a filtering methodology based on machine learning to identify articles that are relevant to the macroeconomic indices in question, and provide a proof of concept demonstrating that emotions expressed in those news items add value to forecasts of industrial production and consumer prices across a diverse range of economies, both in terms of geographic location and size. We complement this with dimensionality reduction and correlation analysis in order to group the more than 600 emotion scores available into a smaller number of interpretable factors. We find emotions associated with ``happiness'' and ``anger'' to yield the highest predictive power across the variables we forecast. To the best of our knowledge, emotions from GDELT's Content Analysis Measure Systems have not yet been used to forecast macroeconomic variables.

\section{Literature review}

This section addresses a selection of existing literature on macroeconomic forecasting with media sentiment.

A rapidly evolving body of literature examines the use of media sentiment and big data for economic forecasting \citep{buono2017big, kapetanios2018big, stern2020network}. The majority of studies forecast economic variables with regression frameworks combining traditional data with positive and negative sentiment classifications based on word count (as opposed to a wider spectrum of emotions). 

Research suggests that positive and negative sentiment from newspaper narrative is an effective tool for monitoring the economic cycle \citep{tuckett2014bringing, shiller2017narrative}. Similarly, newspaper narrative is found to precede a change in economic variables with low frequency shifts correlating well with financial market events. Hence, newspaper narrative can be regarded as a risk management tool \citep{nyman2018news}. 

While most studies focus on a single economy, Baker et al have developed indices of economic uncertainty for a wide range of countries \citep{baker2016measuring, baker2020covid}. They use an autoregressive framework including variables derived from news as well as macroeconomic variables to gauge whether uncertainty shocks foreshadow weaker macroeconomic performance. Findings suggest that effects of policy uncertainty on firms and macro data raises stock price volatility, lowers investment rates and employment growth. Political bias does not significantly impact the uncertainty indices. Nyman and Ormerod \citep{nyman2020text} apply natural language processing techniques to extract uncertainty-related terms from Reuters news and show that they have a causal relationship with the uncertainty index developed by Baker et al \citep{baker2016measuring}.
Fraiberger et al use Reuters news articles to extract positive and negative sentiment \citep{Media_sentiment_intl_asset_prices}. The study finds that news sentiment improves predictions for both developed and emerging equity markets, with global news sentiment linked to sustained foreign investment and thus having a more significant impact on global stock markets than local sentiment.
Thorsrud decomposes unstructured newspaper text into daily news topics and uses them to forecast quarterly GDP growth, producing significantly better predictions compared to central bank forecasts \citep{thorsrud2016nowcasting}. Larson and Thorsrud use indices based on news topics derived from a large Norwegian business newspaper to demonstrate their predictive power for major economic variables as well as asset prices \citep{value_of_ec_news}.
A study by Pekar and Binner demonstrates that adding information on intended purchases from Twitter tweets alongside lagged consumer index values often yields statistically significant improvements over the baseline model that is trained with lag variables alone \citep{pekar2017forecasting}.
Fronzetti Colladon et al build a sentiment index based on the importance of economic keywords in Italian newspapers and show the index's ability to predict Italian stock and bond market volatilities and returns, including during the COVID-19 outbreak in 2020 \citep{Forecasting_fin_mkts_covid}.

Newspaper archives and Twitter are commonly used sour\-ces for raw textual data, however there is a growing body of research using preprocessed sentiment scores. Ortiz combines official statistics with themes from GDELT to track Chinese economic vulnerability in real-time, showing that the index provides valuable insights for policymakers and investors \citep{casanova2017tracking}. Elshendy et al use data from GDELT together with a set of traditional macroeconomic variables and use social network analysis to generate predictors for macroeconomic indices such as consumer confidence, business confidence and GDP for the 10 largest EU economies \citep{elshendy2017big}. Results show that data extracted from GDELT is valuable for predicting macroeconomic variables. Chen examines the effect of the negative narrative in relation to international trade from US presidential candidates in 2016 using average tone from GDELT \citep{chen2019online}. The study concludes that narrative can impact the economy by influencing market participants' expectations. Glaeser et al use reviews from YELP to forecast the local economy. Results from a regression analysis suggest that the data set is a useful complement for predicting contemporaneous changes in the local economy \citep{glaeser2017nowcasting}. YELP data also provides an up-to-date snapshot of economic change at local level, delivering the best results for populous areas and the hospitality industry, given the high number of reviews.

Most publications argue in favour of using media sentiment for macroeconomic forecasting. Schaer et al take a more critical view, highlighting the need for thorough statistical testing, careful choice of error metrics and benchmarks and acknowledging some of the challenges when using sentiment data such as data complexity, sampling instability and key word selection \citep{schaer2019demand}.

The majority of literature only incorporates positive and negative tone to improve macroeconomic predictions, with just a handful of studies featuring a wider range of emotions in their analyses. This paper expands the existing body of research by incorporating nuanced sentiment from newspapers around the world into macroeconomic forecasts of industrial production and consumer prices for 10 diverse economies. This paper goes beyond mere prediction and also focuses on the interpretability of results, illustrating which emotions have the strongest predictive power.

\section{Data and methods}
This section introduces GDELT as data source, outlines the filtering methodology that is used and provides information about the nature of the sentiment scores.

The GDELT Project is a research collaboration of Google Ideas, Google Cloud, Google and Google News, the Yahoo! Fellowship at Georgetown University, BBC Monitoring, the National Academies Keck Futures Program, Reed Elsevier's LexisNexis Group, JSTOR, DTIC and the Internet Archive. The project monitors world media from a multitude of perspectives, identifying and extracting items such as themes, emotions, locations and events. GDELT version two incorporates real-time translation from 65 languages and measures over 2,300 emotions and themes from every news article, updated every 15 minutes \citep{leetaru_the_nodate}. It is a public data set available on the Google Cloud Platform.

The Global Knowledge Graph (GKG), one of the tables within GDELT, contains fields such as sentiment scores and themes extracted from global newspaper articles. It comprises around 11 terabytes of data with new data being added constantly, starting in February 2015. To date, it has analysed over one billion news items.

\subsection{Predicted variables}
This study models industrial production (IP) and consumer price indices (CPI) for US, UK, Germany, Norway, Poland, Turkey, Japan, South Korea, Brazil and Mexico.
IP is a monthly published measure of economic activity. It is defined as the output of industrial establishments, covers a broad range of sectors and tracks the change in the volume of production output. The consumer price index (CPI) is selected as monthly inflation index and describes the change in the prices of a basket of goods and services that are typically purchased by households.

\subsection{Filtering methodology}
\label{sec:filtering}

A filtering methodology is applied to extract sentiment scores from GDELT's GKG relevant to economic growth and inflation, respectively, containing three steps:

\begin{itemize}
  \item Step 1: Keyword filter
  \item Step 2: Classification with neural network
  \item Step 3: Aggregation
\end{itemize}

Step one consists of a top-level thematic filter based on keywords (economic growth, inflation) to select relevant articles based on themes.  Step two uses a neural network to further filter news items using GDELT themes. Step three aggregates the sentiment scores to the frequency of the macro\-economic variables. In addition, this step applies country filters to GDELT locations.

To filter out non-relevant information, a simple keyword filter is applied to GKG themes. The GDELT algorithm extracts themes from every news article it analyses \citep{leetaru2015mining}. The GKG contains over 12,000 unique themes. 

An analysis of a random set of 100 original news articles is conducted to evaluate the keyword filter's ability to eliminate non-relevant news items, showing that the GDELT algorithm has a tendency of recognising themes where there are none. This creates the need to further filter observations using the themes column. 
In GDELT’s GKG, every row corresponds to one analysed news article. The Themes column includes all the themes the GDELT algorithm extracts from a news item as a string of labels, occurring in the same order they are identified in the original text.
A set of 1,000 random articles is manually classified according to relevance (zero not relevant, one relevant). This is done by looking up the original news articles using the DocumentIdentifier column, which corresponds to the article's url. In cases where the url is no longer available, the news item is disregarded. 

Next, the raw GDELT data is preprocessed. Each string of labels is split into lower case tokens. The tokens for every article are then label-encoded so that the themes are given numbers between zero and $N-1$ classes. For out-of-vocabulary words, an ``unknown'' token is assigned. The length for each token sequence is standardized to address the variable length of these sequences by setting a maximum length of 5,000 tokens and padding. 
The encoded themes represent the predictor, and the classification into relevant/non-relevant represent the predicted data, respectively.\\

Model performance is evaluated using k-fold cross-validation as it provides a robust estimate of the performance of a model on unseen data. This is done by dividing the training data set into 10 subsets and taking turns training models on all subsets except one which is held out, and assessing model performance on the held out validation data set. The process is repeated until all subsets are given an opportunity to be the held out validation set. 

Performance is assessed using precision (the number of true positives divided by the number of true positives and false positives), recall (the number of true positives divided by the number of true positives and the number of false negatives) and the F1 score:

\[
    F1 = 2 \times \frac{\mathrm{precision} \times \mathrm{recall}}{\mathrm{precision}+\mathrm{recall}} \ .
\]

He and Ma suggest that these metrics are appropriate in an information retrieval task as they convey the proportion of relevant information identified together with the amount of actually relevant information from the information assessed as relevant by a classifier \citep{he2013imbalanced}. Further, recall and F1 score are  more  suitable  metrics  for  the  assessment  of  a  classifier  than  basic accuracy, especially in the case of imbalanced data, as the latter is too biased towards the dominant class.
Table \ref{tab:performance} shows the performance of different classification algorithms that were evaluated on a data set filtered for economic growth.\\

\FloatBarrier
\begin{center}
\begin{table}[h]
\begin{adjustbox}{width=\columnwidth,center}
\begin{tabular}{|l||*{4}{c|}}\hline

Classifier & Precision & Recall & F1\\
\hline
Gaussian Naïve Bayes & 0.6747 & 0.5000 & 0.5744 \\ 
Random Forest & 0.8304 & 0.9118 &  0.8692\\  
Support Vector Machine & 0.8036 & 0.8645 & 0.8329 \\
Unidirectional NN & 0.8610 & 0.8649 & 0.8571 \\
Bi-LSTM & 0.8853 & 0.9375 &  0.9101\\
\hline
\end{tabular}
\caption{\label{tab:performance}Classifier performance}
\end{adjustbox}
\end{table}
\end{center}
\FloatBarrier

A bidirectional long short term memory (Bi-LSTM) neural network is selected as it exhibits the best performance in terms of Precision and Recall among the different algorithms explored. 
Hochreiter and Schmidhuber introduced long short term memory (LSTM) as a new recurrent neural network (RNN) structure able to span longer time periods without deterioration of short term capabilities \citep{hochreiter1997long}. This architecture imposes constant error flow during backpropagation through internal states of special units and approaches the vanishing gradient issue. This problem is caused by the repeated use of the recurrent weight in RNNs. Therefore, RNNs have difficulties learning long term dependencies.
A LSTM network is a type of RNN that uses special units in addition to standard units, including a ``memory cell'' that can retain information in memory for an extended period. A set of gates controls when information enters the memory, when it is output, and when it is forgotten. This architecture allows LSTM to learn longer-term dependencies. Unidirectional LSTM preserves information from one direction as it only runs forward. A Bi-LSTM runs the inputs in two ways, forwards and backwards. Using the two hidden states combined permits a Bi-LSTM to preserve information from both directions at any point \citep{schuster1997bidirectional}.
Graves and Schmidhuber demonstrate that the Bi-LSTM architecture is well suited to tasks where context is important \citep{graves2005framewise}.

The Bi-LSTM architecture utilised for the classification task contains two hidden LSTM layers, containing 32 and 16 one cell memory blocks, respectively.  The input length is standardised to 5,000. The architecture's details are illustrated in Table \ref{tab:architecture}.

\FloatBarrier
\begin{center}
\begin{table}[h]
\centering
\begin{tabular}{ |c|c|c|c|c| } 
\hline
Layer type & Output shape\\
\hline
Masking & None, 5000 \\ 
Embedding & None, 5000, 16\\  
Bidirectional LSTM & None, 5000, 32  \\
Bidirectional LSTM & None, 5000, 16   \\
Dense & None, 5000, 8\\
Dropout & None, 5000, 8\\
Dense & None, 5000, 1\\
\hline
\end{tabular}
\caption{\label{tab:architecture}Bi-LSTM model architecture}
\end{table}
\end{center}
\FloatBarrier

The ``None'' in the Output shape column indicates no predefined number. By not assigning a specific value, the model has the flexibility to change this number as the batch size changes, inferring the shape from the context of the layers. The masking layer informs the model that some part of the data is padding and should be ignored. The output layer contains a single neuron to make predictions. It uses the sigmoid activation function to produce a probability output in the range of zero to one. In order to map this to a discrete class (zero or one), the threshold of 0.5 is set. Values below this threshold are assigned to the first and values above are assigned to the second class, respectively.

The filtered data from step 2 is aggregated according to time period and location filters are applied to GKG locations according to each country's economic links. The location column contains a list of all locations found in each news item, extracted through the algorithm designed by Leetaru \citep{leetaru2016can}.\\

In order to gain insights into the economic interconnectedness of each of the 10 countries, the import and export volumes by trading partner are examined \citep{oec}. Six of the economies have diversified trade links with countries around the world. Poland, Norway and Turkey trade predominantly with Western European economies. For South Korea, over half of the country's imports and exports are linked to China. Due to the trade links between economies, information relating to one country may also be relevant for another one \citep{piccardi2018complexity}. Based on this idea of interconnection, a global data set incorporating information on all 10 countries is generated for the six global economies (US, UK, Germany, Japan, Brazil, Mexico). For Poland, Norway and Turkey, a data set containing information on Western European economies is generated. For South Korea, a data set including information on China is used.

\subsection{Nature of sentiment scores}
Within GDELT's GKG, the Tone and the Global Content Analysis Measures (GCAM) column contain over 2,300 sentiment scores.

The Tone field comprises a comma-delimited list of six emotional dimensions, each recorded as floating point number. From this field, the average tone of the document is used. This score typically ranges from -10 (very negative) to +10 (very positive), with zero being neutral \citep{noauthor_gdelt_2015-1}. The tone score is based on sentiment mining. This approach counts words according to positive and negative pre-compiled dictionaries. The net sentiment represents the overall tone \citep{hu2004mining}.

The GCAM system runs 24 content analysis systems over each news article and returns the resulting scores as a comma-delimited list into the GCAM column. The majority of GCAM scores is based on word count, some are based on more sophisticated methods. GCAM also includes the overall word count for each news item analysed.

There is some overlap between the GCAM scores generated by the different analysis systems. Scores of the following four analysis systems are chosen as they minimise duplication of sentiment scores while incorporating a broad range of emotions:

\begin{itemize}
\item WordNet-Affect was developed by Strapparava and Valitutti \citep{strapparava2004wordnet} based on WordNet Domains \citep{magnini2000integrating}. WordNet Domains maps synsets, i.e. groupings of synonymous words expressing the same concept, to domain labels such as Economics or Health. WordNet-Affect extends this structure in assigning affective domain labels to the synsets. WordNet-Affect scores are word count-based and account for 280 sentiment dimensions such as ``joy'', ``fear'' or ``sadness'' in the GCAM column. For example, appearances of the word ``joy'' in the original text will increase the ``joy'' score.
\item The Loughran and McDonald Financial Sentiment Dictionary uses negative word lists specific to a financial context to produce scores based on word count. The authors find that word lists for other disciplines often misclassify words in financial documents \citep{loughran2011liability}. For example, words such as tax, cost, capital, and liability are typically not negative in a financial context but have a negative association in the Harvard dictionary. The system generates six scores. 
\item The Hedonometer scores provide a measurement for overall societal happiness for English and a range of non-English languages \citep{dodds2011temporal}. In order to provide an overall score, over 10,000 unique words are rated by humans on a scale from one to nine. For each of these words, an average happiness score is derived, with five being neutral. For instance, laughter, food and hate have been rated 8.5, 7.4 and 2.3. To derive the happiness score of a text, the average happiness level is calculated. The system returns 12 scores.
\item ML-Senticon represents a multi-layered synset-level lexicon and calculates positivity and negativity scores covering English and Spanish \citep{cruz2014building}.  First, the synsets are assigned polarity scores. Then, these scores are fine-tuned by creating a graph of synsets, where the nodes represent synsets. Edges between nodes exist if synset i is in synset j. Lastly, a type of random-walk algorithm propagates the positivity (negativity) scores from the previous step through the edges of the graph to derive the positivity (negativity) values for each synset. ML-SENTICON groups synsets (sets of words with the same meaning) into eight successive layers, with each layer adding more but lower-confidence synsets, allowing to tune for recall versus precision in scoring those synsets ``positive'' or ``negative''. The system provides 32 scores.
\end{itemize}
See appendix \ref{app:sentiment} for further details on the above sentiment scores.

The extracted data is aggregated by month. The mean and standard deviation of the tone score is calculated. Where the GCAM sentiment scores are based on word count, the mean and standard deviation are calculated, normalized to account for variation of word count as done by Baker et al. \citep{baker2016measuring}. For calculated sentiment scores, the mean score and standard deviation over the period are computed. In addition, the number of news items and the total word count per period is generated. 

\subsection{The data sets}
This section sets out how the data sets used in this study are created.

The filtering methodology is applied to build two data sets, filtered for articles relevant to economic growth and inflation, respectively and country filters for the US, UK, Germany, Norway, Poland, Turkey, Japan, South Korea, Brazil and Mexico are applied. The choice of countries used in the analysis reflects an even split in developed and developing countries as per the MSCI Emerging Market Index \citep{MSCI}. For both groups, we select a diverse mix of economies, both in size, drivers of economic growth and geography. For instance, Germany is a large developed eurozone country whose economy is export-driven (cars, machinery), with diversified economic links. Turkey is classified as a developing economy that exports the majority of its goods (agricultural produce, textiles, steel) to European countries. A further criterion is the availability of reliable macroeconomic data for all of the selected countries.

The data is aggregated to monthly frequency, from beginning of March 2015 to end of June 2020, respectively.\
Model predictions incorporate true positives, false positives, true negatives and false negatives. The filtered data sets comprise true positive and false positive predictions only, which corresponds to c 5.4\% and c 3.9\% of noise for the economic growth and the inflation filter, respectively.

An unfiltered data sample of aggregated GCAM scores is created for comparison purposes. Around five million random observations (one million for each calendar year) are selected and aggregated to monthly frequency. The unfiltered data set contains over 60\% noise i.e. news items not relevant to economic growth or inflation, respectively.\\

In order to account for macroeconomic effects, the baltic dry index and the crude oil price are incorporated when modelling IP. The baltic dry index is a leading indicator for economic activity, reflecting levels of global trade \citep{bildirici2015baltic}. A study by Van Eyden et al suggests that there is a significant relationship between oil price fluctuations and economic growth in OECD countries \citep{van2019oil}. For models forecasting CPI, the countries' respective terms of trade indices as well as the crude oil price are included. Mihailov et al find that the anticipated relative change in the terms of trade is a more important determinant of inflation than the contemporaneous domestic output gap \citep{mihailov2011small}. A study by Salisu et al establishes a significant long-term positive relationship between oil price and inflation \citep{salisu2017modelling}.

\subsection{Data preprocessing}
In this section the data preparation methods are summarized.

For each of the 10 aforementioned economies the respective values for IP and CPI are used as predicted variables. Both index values represent the monthly percentage change.

The augmented Dickey Fuller unit root test is applied to 20 years of monthly data and stationarity is not rejected at 5\% significance for the above described variables.

Where a sentiment score contains zeros only, it is assumed that the relevant GCAM system did not return any scores and they are dropped from the respective data set. The scores affected are mainly based on the Hedonometer and ML Senticon GCAM systems. The GDELT data sets contain 664 raw scores and this step reduces the amount of features to 630 and 628 for data sets filtered for economic growth and inflation, respectively. The unfiltered data set retains 632 features. The monthly change in sentiment scores is applied. The augmented Dickey Fuller unit root test is applied and stationary is not rejected at 5\% for any of the scores. The sentiment scores are  standardized by removing the mean and scaling to unit variance.

\section{Analysis}
This section outlines the analysis that is performed to gauge if the sentiment scores from GDELT have predictive power.

\subsection{Granger causality analysis}
The Granger causality between the GDELT sentiment scores and the  predicted variables is assessed to evaluate if there are relationships between those variables. While Granger causality can provide useful insights into the relation between variables, it is not testing true causality, instead, the test looks to establish if changes in one variable occur before changes in the other one \citep{granger1969investigating}. This means that Granger causality may be found even when there is no causal link \citep{leamer_1985}.

The null hypothesis for the Granger causality test states that lagged sentiment scores are not causing a variable at a significance level of 5\%, while the alternate hypothesis stipulates that lagged sentiment scores are Granger-causing an index at the same significance level.

The Granger causality for lags up to a maximum of three months is evaluated. Since multiple tests for each data set are run, the resulting $p$-values are adjusted according to the Benjamini-Hochberg (BH) procedure to control for multiple hypothesis testing \citep{benjamini2005false}.

\subsection{Forecasting} 
As a further step in the analysis of the sentiment scores, a three step approach is used for forecasting the macroeconomic variables as proposed by Girardi, Guardabascio and Ventura \citep{girardi2016factor}.

In a first step, an autoregressive model including the predicted variable and the explanatory macroeconomic variables only is used to predict industrial production and inflation, respectively.

This framework allows modeling a \(T \times K\) multivariate time series \(Y\), where \(T\) denotes the number of observations and \(K\) the number of variables. The framework is defined as
\begin{equation}\label{eq:1}
  Y_{t } =  v + A_{1}Y_{t-1}+\dots +  A_{p}Y_{t-p} + u_{t}
\end{equation}
where \(A_{i}\) is a \(K \times K\) coefficient matrix, $v$ is a constant and $u_{t}$ is white noise.\\

Due to the large number of GDELT scores, they cannot be incorporated in the autoregressive framework described in Eq. \ref{eq:1}. Therefore, as a second step, factors are extracted from the broad set of GDELT sentiment scores for inclusion into the framework.
Partial Least Squares (PLS) is applied as dimensionality reduction technique to obtain useful information from the GDELT scores. This technique is appropriate where the number of features is significantly larger than the number of observations, and features are correlated \citep{cubadda2012medium}. PLS includes information from predicted variable and predictors when deriving scores and loadings, which are selected to maximise the covariance between predicted variable and predictors \citep{de1993simpls}.

PLS is implemented on the residuals derived at step one. The residuals include the portion of the predicted variable that is not explained and therefore, applying PLS to the GDELT sentiment scores provides additional information to the predictors. The orthogonal relationship between the predicted variable and the residuals maintains the orthogonality between the factors extracted by PLS and the autoregressive components.
For both IP and CPI, the first three PLS components account for around 80\% of the variation in the predicted variables. Cross-validation analysis shows that the residual sum of squares are increasing in a model with more than three factors, indicating that three PLS factors are appropriate \citep{tobias1995introduction} (see results for US variables in Table \ref{tab:pls}).

\FloatBarrier
\begin{center}
\begin{table}[h]
\begin{adjustbox}{width=\columnwidth,center}
\begin{tabular}{|l||*{4}{c|}}\hline
No of factors    & IP: R \textsuperscript{2} & IP: RSS & CPI: R \textsuperscript{2} & CPI: RSS \\
\hline
2   & 0.6733 & 137.5931 & 0.7496 & 56.0192\\
3   & 0.7755 & 64.4186 & 0.8365 & 32.1136\\
4  & 0.7876 & 65.2438 & 0.8573 & 45.5851\\
5  & 0.7880 & 66.7896 & 0.8601 & 50.8778\\\hline

\hline
\end{tabular}
\caption{\label{tab:pls}Results from PLS regression analysis on US IP and US CPI for different numbers of factors.}
\end{adjustbox}
\end{table}
\end{center}
\FloatBarrier

As a third step, for each country, the respective predicted variable, the respective explanatory variables and the three PLS components derived from the GDELT sentiment are used as input into the autoregressive framework described in Eq. \eqref{eq:1} to form a factor augmented autoregressive model \citep{colladon2019using}. 

The model is then calibrated for each macroeconomic variable and each country.

The optimal lag length is selected based on the the Akaike (AIC) and the Bayesian (BIC) information critera. These measures are based on the idea that the inclusion of a further term may improve the model however the model should also be penalised for increasing the number of parameters to be estimated. When the improvement in goodness-of-fit outweighs the penalty term, the statistic associated with the information criterion decreases. Thus, the lag which minimises the information criterion is selected \citep{book}.

Three benchmarks are used to compare model performance -- first, an autoregressive framework including the predicted variable and explanatory macroeconomic variables, second, an autoregressive framework incorporating the predicted variable, explanatory macroeconomic variables and unfiltered GDELT sentiment factors and third, an autoregressive framework incorporating the predicted variable, explanatory macroeconomic variables and the average tone score from GDELT. 

Performance is assessed using walk-forward cross-validation and the root mean squared error (RMSE). The data set is split into three folds. This cross-validation technique is suitable for time series data as in the k\textsuperscript{th} split, it returns the first k folds as train set and the (k+1)\textsuperscript{th} fold as test set.

The modified Diebold Mariano test proposed by Harvey, Leybourne and Newbold \citep{harvey1997testing} is used to gauge whether model forecasts are significantly different. 

\section{Research findings}
This section presents the findings from the analysis set out in the previous section.

\subsection{Granger causality test results}
The sentiment scores from GDELT and the macroeconomic indices for 10 countries are tested for Granger causality, with a maximum lag of three months. Tables \ref{tab:ip_granger} and \ref{tab:cpi_granger} display the number of BH-adjusted $p$-values that exhibit significance at 5\% for each country's macroeconomic variable. The ``Filtered'' column refers to results from models including GDELT sentiment scores filtered for economic growth and inflation respectively, while the ``Unfiltered'' column shows the results for the models incorporating the unfiltered GDELT sentiment.

\FloatBarrier

\begin{center}
\begin{table}[h]
\begin{tabular}{|l||*{3}{c|}}\hline
\backslashbox{Country}{Data set}
&\makebox[6em]{Filtered}&\makebox[6em]{Unfiltered}\\\hline
US & 5 & 0\\  
UK & 29 & 0\\
Germany & 8 & 7\\
Norway & 30 & 0\\
Poland & 12 & 0\\
Turkey & 8 & 1\\
Japan & 6 & 0\\
South Korea & 10 & 0\\
Brazil & 35 & 16\\
Mexico & 12 & 0\\
\hline
\end{tabular}
\caption{IP: Number of significant BH-adjusted $p$-values}
\label{tab:ip_granger}
\end{table}
\end{center}

\begin{center}
\begin{table}[h]
\begin{tabular}{|l||*{2}{c|}}\hline 
\backslashbox{Country}{Data set}
&\makebox[6em]{Filtered}&\makebox[6em]{Unfiltered}\\\hline
US & 14 & 0\\  
UK & 30 & 8\\
Germany & 13 & 3\\
Norway & 11 & 0\\
Poland & 16 & 3\\
Turkey & 57 & 1\\
Japan & 19 & 0\\
South Korea & 17 & 0\\
Brazil & 39 & 0\\
Mexico & 35 & 15\\
\hline
\end{tabular}
\caption{CPI: Number of significant BH-adjusted $p$-values}
\label{tab:cpi_granger}
\end{table}
\end{center}

\FloatBarrier

Notwithstanding the limitations of the Granger causality test \citep{leamer_1985}, the results show a pattern. For both macroeconomic variables, the filtered data sets exhibit consistent Granger causality across countries.

The analysis suggests that the filtering methodology introduced in section~\ref{sec:filtering} adds value and is able to generate sentiment scores that have a relationship with economic indices. 

Some reverse Granger causality exists between macroeconomic variables and GDELT sentiment scores albeit much sparser than that shown in \ref{tab:ip_granger} and \ref{tab:cpi_granger}. Therefore, there is more consistent evidence that sentiment from news Granger causes the macroeconomic variables considered than vice-versa.

\subsection{Forecast error analysis}

The respective filtered sentiment data sets are condensed into three components using PLS. They are then used to predict IP and CPI, for 10 countries each with the model in Eq. (\ref{eq:1}). All models have a lag of one month, determined by evaluating AIC and BIC.

The columns of Tables \ref{tab:IP_results} and \ref{tab:CPI_results} show the performance of the forecasts from models containing filtered GDELT sentiment factors compared to three benchmarks, which consist of models only including predicted variable and explanatory macroeconomic variables (referred to as BM1), predicted variable, explanatory macroeconomic variables and unfiltered GDELT sentiment factors (referred to as BM2) and models including predicted variable, explanatory macroeconomic variables and the average tone score from GDELT. (referred to as BM3).  The numbers in the cells represent the RMSE in percentage terms for each model and its benchmarks. Blue (red) cells denote cases in which the models outperform (underperform) the respective benchmarks. In the column ``Sign.'', numbers in parentheses correspond to the number of significant coefficients associated with GDELT factors in the model in Eq. (\ref{eq:1}), with the asterisks denoting the level of their statistical significance.
For example, the first row in Table \ref{tab:IP_results} conveys the fact that the model containing the filtered US GDELT factors outperforms all three benchmarks, with all three factors being statistically significant (at 0.1, 0.05 and 0.01, respectively). In the case of Norway, instead, the model containing filtered factors only outperforms the second benchmark model, with only one statistically significant factor.

\FloatBarrier
\begin{center}
\begin{table}[h]
\begin{adjustbox}{width=\columnwidth,center}
\begin{tabular}{|l||*{5}{c|}}\hline
\backslashbox{IP for}{Data set}
 &\makebox[4em]{Model}&\makebox[4em]{BM1}& \makebox[4em]{BM2}& \makebox[4em]{BM3}& \makebox[4em]{Sign.}\\\hline
US & 1.6348 & \cellcolor{blue!15} 1.6439& \cellcolor{blue!15}1.6527& \cellcolor{blue!15}1.6507&***(1),**(1), *(1)\\
UK & 2.4663 & \cellcolor{blue!15}2.4870& \cellcolor{blue!15}2.5065& \cellcolor{blue!15}2.4887& ***(1)\\
Germany & 2.7957 & \cellcolor{blue!15} 2.7978& \cellcolor{red!15}2.7870& \cellcolor{blue!15}2.8032&**(2)\\
Norway & 2.2907& \cellcolor{red!15}2.2647 & \cellcolor{blue!15}2.3410& \cellcolor{red!15}2.2720&**(1)\\
Poland & 7.6308 & \cellcolor{blue!15}7.7321 & \cellcolor{blue!15}7.8420& \cellcolor{blue!15}7.7351&***(1)\\
Turkey & 4.5785 & \cellcolor{blue!15}4.5836 & \cellcolor{blue!15}4.6780& \cellcolor{blue!15}4.5843&**(2)\\
Japan & 2.2138& \cellcolor{blue!15} 2.2429& \cellcolor{blue!15}2.2571& \cellcolor{blue!15}2.2429&***(1), **(1)\\
South Korea & 2.4776& \cellcolor{blue!15}2.5079 & \cellcolor{blue!15}2.5579 & \cellcolor{blue!15}2.5211&***(1)\\
Brazil & 4.0484 & \cellcolor{blue!15}4.0942 & \cellcolor{blue!15} 4.1341& \cellcolor{blue!15}4.0941&**(1)\\
Mexico & 3.2630 & \cellcolor{red!15} 3.2494& \cellcolor{red!15}3.2532& \cellcolor{red!15}3.2471&***(1)\\\hline

\hline
\end{tabular}
\caption{Results of the model in Eq. \ref{eq:1} applied to IP. \newline Numbers represent the RMSE (\%). Blue (red) cells denote cases in which the model outperforms (underperforms) the benchmark. Numbers in parentheses correspond to the number of significant coefficients associated with GDELT factors in the model in Eq. (\ref{eq:1}) (${***}$ denotes at least one GDELT sentiment factor with $p$-value < 0.01, ${**}$ $<$ 0.05, ${*}$ $<$ 0.1).}
\label{tab:IP_results}
\end{adjustbox}
\end{table}
\end{center}

\FloatBarrier

The models including filtered GDELT sentiment factors outperforms the benchmarks in eight out of ten cases. All of these models contain one or more statistically significant GDELT sentiment factors.
The modified Diebold Mariano test shows that model forecasts for IP are different from BM1 and BM2 in nine out of ten, and for BM3 in 10 cases, respectively at either 1, 5 or 10\% significance (see Table \ref{tab:DM_IP_results}).\\

During the first half of 2020, IP for all ten countries experienced high levels of volatility as governments around the world imposed lockdowns that severely impacted economic activity. In the cross-validation, the last validation set incorporates the period of the COVID-19 outbreak in 2020. Predictions on this last validation set show a much larger error metric across countries than those predictions on validation sets that exclude the outbreak. However, performance dynamics during the COVID-19 outbreak remain the same in that the model including GDELT factors outperforms the three benchmarks for most countries.\\

\FloatBarrier

\begin{center}
\begin{table}[h]
\begin{adjustbox}{width=\columnwidth,center}
\begin{tabular}{|l||*{5}{c|}}\hline
\backslashbox{CPI for}{Data set}
&\makebox[4em]{Model}&\makebox[4em]{BM1}& \makebox[4em]{BM2}& \makebox[4em]{BM3}& \makebox[4em]{Sign.}\\\hline
US & 0.2051 & \cellcolor{red!15} 0.2031& \cellcolor{blue!15}0.2165& \cellcolor{red!15}0.2038&***(1)\\
UK & 0.3212 & \cellcolor{red!15}0.3118& \cellcolor{blue!15}0.3258& \cellcolor{red!15}0.3193& ***(2)\\
Germany & 0.4117 & \cellcolor{blue!15}0.4316 & \cellcolor{blue!15}0.4288& \cellcolor{blue!15}0.4279&***(1)\\
Norway & 0.4715 & \cellcolor{blue!15}0.4727 & \cellcolor{red!15}0.4706& \cellcolor{red!15}0.4640&\\
Poland & 0.3321 & \cellcolor{blue!15} 0.3383& \cellcolor{blue!15}0.3649& \cellcolor{blue!15}0.3512&***(1)\\
Turkey & 1.0194 & \cellcolor{blue!15}1.0263 & \cellcolor{blue!15}1.0513& \cellcolor{blue!15}1.0685&***(2), **(1)\\
Japan & 0.2464 & \cellcolor{blue!15}0.2485 & \cellcolor{blue!15}0.2538& \cellcolor{blue!15}0.2524&***(2)\\
South Korea & 0.3941 & \cellcolor{blue!15}0.4055 & \cellcolor{blue!15}0.4057 & \cellcolor{blue!15}0.4070&**(1)\\
Brazil & 0.3876 & \cellcolor{blue!15} 0.3993& \cellcolor{blue!15}0.4165 & \cellcolor{blue!15}0.3946&***(1)\\
Mexico & 0.4226 & \cellcolor{blue!15}0.4349 & \cellcolor{blue!15}0.4340& \cellcolor{blue!15}0.4581&***(1)\\\hline

\hline
\end{tabular}
\caption{Results of the model in Eq. \ref{eq:1} applied to CPI. \newline Numbers represent the RMSE (\%). Blue (red) cells denote cases in which the model outperforms (underperforms) the benchmark. Numbers in parentheses correspond to the number of significant coefficients associated with GDELT factors in the model in Eq. (\ref{eq:1}) (${***}$ denotes at least one GDELT sentiment factor with $p$-value < 0.01, ${**}$ $<$ 0.05, ${*}$ $<$ 0.1).}
\label{tab:CPI_results}
\end{adjustbox}
\end{table}
\end{center}

\FloatBarrier

The models including filtered GDELT sentiment factors outperform BM1 for eight, BM2 for nine and BM3 for seven out of ten countries, respectively. Nine out of ten models contain one or more statistically significant GDELT sentiment factors.
According to the adjusted Diebold Mariano test, model forecasts for CPI are different from BM1 for all and for BM2 and BM3 for seven out of ten countries, respectively at either 1, 5 or 10\% significance (see Table \ref{tab:DM_CPI_results}).\\

Results suggest that the filtering methodology isolates relevant signals, given that those models using filtered sentiment perform consistently better that those using unfiltered sentiment. Further, the findings indicate that sentiment scores extracted from GDELT improve predictions for IP and CPI for most countries. Results show that incorporating a wide spectrum of emotions in forecasts yields better forecasts compared to only including one sentiment score such as the average tone.

\subsection{Drivers of GDELT factors}
In order to gain insights into the relationship between sentiment scores and the PLS components derived from the filtered GDELT data, the loadings corresponding to each component are examined. 

Loadings correspond to the strength of relationship between the original sentiment scores and the PLS components, quantifying the relevance of the underlying sentiment scores in each of the components. 

All sentiment scores from GDELT represent a specific emotion such as ``cheerfulness'', ``euphoria'' or ``joy'' and are manually mapped to seven universal emotions as set out by Ekman and Corduro \citep{ekman2011meant}. These seven emotions define emotions as discrete, automatic reactions to events and stipulate that emotions such as happiness or anger describe groups of related states with distinct common traits. According to these seven groups, the above mentioned emotions are assigned to ``happiness''.

For each component, the loadings are summed according to these seven distinct emotions. Mapping sentiment scores onto emotions provides some interpretability to our analysis, by allowing us to investigate which emotions are associated with each PLS component.

\FloatBarrier
\begin{figure}[pos=h]
	\centering
	\includegraphics[scale=.6]{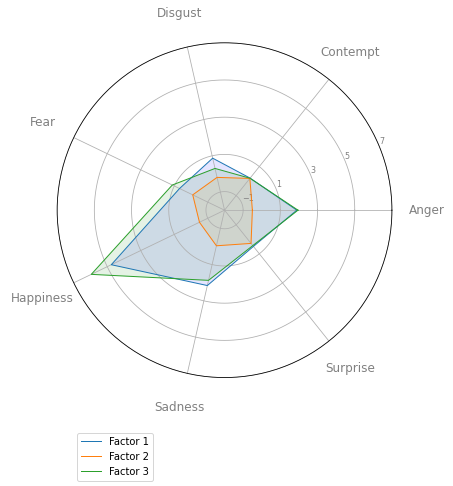}
	\caption{IP: Significant PLS components explained by emotions (US)}
	\label{FIG:2}
\end{figure}
\FloatBarrier

\FloatBarrier
\begin{figure}[pos=h]
	\centering
	\includegraphics[scale=.6]{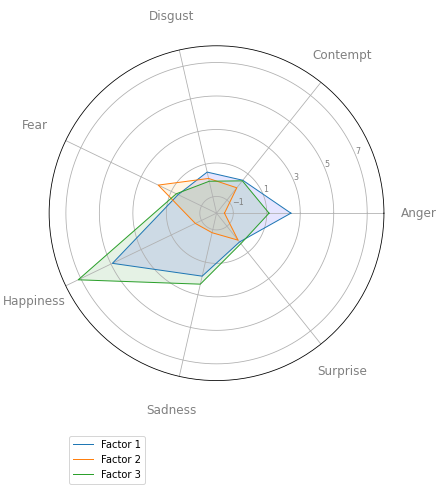}
	\caption{CPI: Significant PLS components explained by emotions (Turkey)}
	\label{FIG:3}
\end{figure}
\FloatBarrier

As an example, in Figs. \ref{FIG:2} and \ref{FIG:3} we show radar charts of the emotions associated to the loadings corresponding to the statistically significant PLS components used to forecast IP and CPI in the US and Turkey, respectively. As can be seen from Tables \ref{tab:IP_results} and \ref{tab:CPI_results}, the corresponding models outperform the benchmarks we considered and are associated with substantial statistical significance. The components explain 78\% and 84\% of the variation in the components shown in Figs. \ref{FIG:2} and \ref{FIG:3}, respectively.
Further charts can be provided upon request.

The findings from this example show that the factors we use to predict IP and CPI can be associated with well defined emotions. Therefore, movements in such emotions -- as expressed in news articles published by global newspapers -- contribute to explain movements in major macroeconomic indices. Of the seven distinct emotions, ``happiness'' and ``anger'' have the strongest predictive power, both with positive relationships. This is the case across all PLS components. 

\section{Discussion}
This study proposes a new method of incorporating emotions from global newspaper articles into macroeconomic forecasts.
It introduces a filtering methodology to extract and aggregate large volumes of data. The methodology is applied to build data sets filtered for economic growth and inflation. The country-specific macroeconomic indices are forecast using data sets for IP and CPI, respectively, that take into account each country's trade links when applying location filters. The filtered data exhibits consistent Granger causality across the two macroeconomic variables. Autoregressive models including the filtered data outperform their benchmarks for most predicted variables. In particular, models incorporating a wide spectrum of emotions mostly outperform those models only including one aggregate sentiment score, average tone. Mapping the GDELT sentiment scores onto distinct emotions helps understand how these emotions relate to each PLS component and thus interpret our analysis, suggesting that ``happiness'' and ``anger'' are their main drivers. 

Our work advances the literature on macroeconomic forecasting with news data in at least two main respects. First, our use of sentiment scores is -- to the best of our knowledge -- new. Indeed, our approach leverages four distinct sentiment analysis methodologies, and synthesises them into a small number of factors. This is somewhat akin to ensemble approaches in machine learning, which seek to combine a variety of models into one better performing ``meta-model''. In contrast, the majority of other approaches to sentiment-based macroeconomic forecasting only leverage simple tone-based sentiment analysis methods \citep{chen2019online,glaeser2017nowcasting,casanova2017tracking}. Second, the factors we extract lend themselves to a fairly intuitive interpretation in terms of major groups of emotions (see Figs. \ref{FIG:2}) and \ref{FIG:3}. Such factors represent echoes of the ``animal spirits'' \citep{keynes2018general} and ``visceral factors'' \citep{loewenstein2000emotions} that drive human economic behaviour.  In this respect, our work provides a data-driven framework to operationalise such concepts and to incorporate them into quantitative predictive economic models.

\subsection{Limitations and ideas for further research}
Before concluding, we ought to acknowledge a few potential limitations of our study. First, we only examine linear relationships between the factors we extract from GDELT data and macroeconomic indices. Investigating non-linear interactions between these variables could potentially generate further insights and could be an extension to this experiment. Second, our study is a proof of concept and does not attempt to fully optimise performance. Furthermore, it only focuses on forecasting two macroeconomic variables, while only employing a limited selection of other macro indicators as controls. Further work to improve real-world applicability could be done by expanding to a much bigger array of indicators. In this respect, our approach could be easily transferred to other data-rich domains with an established literature on quantitative forecasting, such as financial markets. Third, GDELT data starts at the end of February 2015 and thus has a short track record. Particularly when modelling monthly data, the small amount of observations is likely to impact the significance of results. There are currently plans to backfill GDELT with additional data going back to 1979. When this data will become available, it will be interesting to repeat this experiment.

\section{Conclusions}
This study introduces a new method of incorporating newspaper sentiment into macroeconomic forecasts. It expands the existing body of research on forecasting  macroeconomic variables by incorporating a wide array of emotions from newspapers around the world. To the best of our knowledge, the GCAM sentiment scores from the GDELT GKG have not yet been used to forecast macroeconomic variables; hence the experiment introduces a new data source.

The study represents a proof of concept showing that the filtering methodology presented captures relevant signals and that the data extracted from GDELT adds value when forecasting macroeconomic variables. The findings demonstrate that the sentiment factors derived from GDELT we use to predict IP and CPI can be linked to distinct emotions. Therefore, fluctuations in such emotions – as expressed in news articles published by global newspapers – help explain changes in major macroeconomic indices.

\section{Appendix}

\subsection{GCAM sentiment scores}
\label{app:sentiment}
This section provides an overview of the GCAM sentiment scores used in this study.\\

\textbf{ML Senticon}
\begin{itemize}
  \item Level 1 to Level 8 Positive (Spanish)
  \item Level 1 to Level 8 Negative (Spanish)
   \item Level 1 to Level 8 Positive (English)
  \item Level 1 to Level 8 Negative (English)
\end{itemize}

\textbf{Hedonometer}
\begin{itemize}
  \item Happiness (English)
  \item Happiness (French)
   \item Happiness (German)
   \item Happiness (Spanish)
  \item Happiness (Hindu)
  \item Happiness (Indonesian)
  \item Happiness (Korean)
  \item Happiness (Arabic)
  \item Happiness (Portuguese)
  \item Happiness (Russian)
  \item Happiness (Urdu)
  \item Happiness (Chinese)
\end{itemize}

\textbf{Loughran \& McDonald Financial Dictionary}
  \begin{itemize}
  \item Litigious
  \item ModalStrong
   \item ModalWeak
   \item Negative
  \item Positive
  \item Uncertainty
\end{itemize}

\textbf{WordNet-Affect}\\
Due to the large number of WordNet-Affect scores, we list a subset for illustrative purposes.
  \begin{itemize}
    \item Abashment
  \item Abhorrence
  \item Admiration
  \item ...
  \item world-weariness
  \item worship
  \item wrath
\end{itemize}

\subsection{P-values from modified Diebold Mariano test}

Tables \ref{tab:DM_IP_results} and \ref{tab:DM_CPI_results} show the p-values from the modified Diebold Mariano test. This test is used to to gauge if model forecasts containing filtered GDELT sentiment factors are significantly different from the forecasts derived from the benchmark models as set out in section 4.2.

\FloatBarrier

\begin{center}
\begin{table}[h]
\begin{adjustbox}{width=\columnwidth,center}
\begin{tabular}{|l||*{4}{c|}}\hline
\backslashbox{IP for}{Data set}
&\makebox[6em]{Model - BM1}& \makebox[6em]{Model - BM2}& \makebox[6em]{Model - BM3}\\\hline
US & 0.0000 & 0.0003 & 0.0000\\
UK & 0.1848 &  0.1007 & 0.0812 \\
Germany & 0.0103 & 0.0093 & 0.0091\\
Norway & 0.0178 & 0.1081 & 0.0181\\
Poland & 0.0338 & 0.0094 & 0.0471\\
Turkey & 0.1014 & 0.0017 & 0.1014\\
Japan & 0.0025 & 0.0812 & 0.0025 \\
South Korea & 0.0450 & 0.9890 & 0.0600 \\
Brazil & 0.0052 & 0.0865 & 0.0053\\
Mexico & 0.0767 & 0.0472 & 0.1009\\\hline

\hline
\end{tabular}
\caption{P-values from modified Diebold Mariano test (IP)}
\label{tab:DM_IP_results}
\end{adjustbox}
\end{table}
\end{center}

\FloatBarrier

The modified Diebold Mariano test shows that model forecasts for IP are different from BM1 and BM2 in nine out of ten, and for BM3 in 10 cases, respectively at either 1, 5 or 10\% significance.

\FloatBarrier

\begin{center}
\begin{table}[h]
\begin{adjustbox}{width=\columnwidth,center}
\begin{tabular}{|l||*{4}{c|}}\hline
\backslashbox{CPI for}{Data set}
&\makebox[6em]{Model - BM1}& \makebox[6em]{Model - BM2}& \makebox[6em]{Model - BM3}\\\hline
US & 0.0132 & 0.4314 & 0.9216\\
UK & 0.0238 & 0.1730 & 0.0349\\
Germany & 0.0701 & 0.1014 & 0.0013\\
Norway & 0.0111 & 0.0599 & 0.0091\\
Poland & 0.0072 & 0.0789 & 0.0027\\
Turkey & 0.0169 & 0.0017 & 0.0010\\
Japan & 0.0152 & 0.0812 & 0.0342 \\
South Korea & 0.0161 & 0.0345 & 0.2579 \\
Brazil & 0.0029 & 0.0481 & 0.0599\\
Mexico & 0.0028 & 0.8028 & 0.9100\\\hline

\hline
\end{tabular}
\caption{P-values from modified Diebold Mariano test (CPI)}
\label{tab:DM_CPI_results}
\end{adjustbox}
\end{table}
\end{center}

\FloatBarrier

Model forecasts for CPI are different from BM1 for all countries and for BM2 and BM3 for seven out of ten countries, respectively at either 1, 5 or 10\% significance.

\section*{Declaration of competing interest}
The authors declare that they have no known competing financial interests or personal relationships that could have appeared to influence the work reported in this paper.

\section*{Acknowledgments}
GL acknowledges support from an EPSRC Early Career Fellowship [Grant No. EP/N006062/1]. The authors thank Simone Righi and David Tuckett for very helpful feedback on preliminary versions of our manuscript.

\bibliographystyle{apacite}

\bibliography{cas-refs}

\end{document}